\documentclass[twocolumn,showpacs,preprintnumbers,amsmath,amssymb]{revtex4}

\usepackage{graphicx}
\usepackage{dcolumn}
\usepackage{bm}
\def\bra#1{\left\langle #1 \right|}
\def\ket#1{\left| #1 \right \rangle}
\def\Be{^{9}\mathrm{Be}^{+}}
\def\Mg{^{24}\mathrm{Mg}^{+}}
\begin{document}

\title{Sympathetic cooling of $\Be$ and $\Mg$ for quantum logic}

\author{M. D. Barrett}
\author{B. DeMarco}
\author{T. Schaetz}
\author{D. Leibfried}
\author{J. Britton}
\author{J. Chiaverini}
\author{W. M. Itano}
\author{B. Jelenkovi\'{c}}
\altaffiliation[Present Address: ]{Institute of Physics, Belgrade,
Yugoslavia}
\author{J. D. Jost}
\author{C. Langer}
\author{T. Rosenband}
\author{D. J. Wineland}
\address{NIST Boulder, Time and Frequency Division, Ion Storage Group}
\date{\today}

\begin{abstract}
We demonstrate the cooling of a two species ion crystal consisting
of one $\Be$ and one $\Mg$ ion.  Since the respective cooling
transitions of these two species are separated by more than
$30\:\mathrm{nm}$, laser manipulation of one ion has negligible
effect on the other even when the ions are not individually
addressed. As such this is a useful system for re-initializing the
motional state in an ion trap quantum computer without affecting
the qubit information. Additionally, we have found that the mass
difference between ions enables a novel method for detecting and
subsequently eliminating the effects of radio frequency (RF)
micro-motion.
\end{abstract}
\maketitle

\section{Introduction}

A promising system for the development of a quantum computer is a
collection of cold trapped ions.  In this scheme, information is
stored in the internal states of the ions, and logic gates are
performed by coupling qubits through a motional degree of freedom.
The original proposal by Cirac and Zoller \cite{gate1,added2}
requires the system to be initialized in the motional ground
state, with imperfect ground-state occupation resulting in a loss
of gate fidelity. Other gate implementations that relax this
condition have been proposed
\cite{gate2,gate3,gate5,gate6,gate7,gate8} and demonstrated
\cite{gate9,gate4} but many of these schemes
\cite{gate2,gate3,gate5,gate8} require the ions to be in the
Lamb-Dicke limit \cite{bible}. Maintaining this condition in a
large scale device places stringent requirements on allowable
heating rates. Furthermore, one proposed architecture for a
large-scale device \cite{bible,Kielpinski} requires separation and
shuttling of ions between different trapping regions in an array
of interconnected traps and initial experiments reported
substantial heating during the separation process \cite{Mary}. For
these reasons it is expected that cooling between gate operations
will be needed in a viable large scale processor.

Ion cooling is typically achieved by laser cooling, which requires
internal state relaxation.  Therefore, direct laser cooling of the
qubit ions is not possible without destroying the coherence of the
qubit state.  Alternatively, additional refrigerant ions can be
laser cooled directly, allowing the motional degrees of freedom to
be sympathetically cooled via the Coulomb interaction
\cite{added1}. For this strategy to work, the cooling radiation
must not couple to the qubit's internal state. Thus the cooling
radiation must be sufficiently focused onto the refrigerants
and/or be sufficiently detuned from transitions within the qubit's
internal state manifold. In the context of quantum computing,
there have been two previous implementations of this scheme. In
one approach the refrigerant ions were of the same species as the
qubit ions, and successful sympathetic cooling hinged on the
ability to individually address the trapped ions \cite{cooling1}.
While this implementation was able to achieve $99\%$ ground state
cooling, individual addressing can be technically demanding,
particularly in tightly confining traps, which are required to
maintain fast gate speeds \cite{bible,Steane}.  In a second
approach the refrigerant ions were differing isotopes of the same
atomic species \cite{cooling2}.  Since isotope shifts of the
relevant cooling transitions are typically several gigahertz and
much larger than natural linewidth, the required degree of
individual addressing is reduced.  Furthermore, the isotope shifts
can be small enough so that optical modulators can provide the
cooling light without the need for additional lasers
\cite{cooling2}. However, not all atomic species have isotopes
that would be suitable for sympathetic cooling. Moreover, the
amount of sympathetic cooling required may require cooling
transitions that are much further detuned from any transitions in
the qubit ion. For these reasons we have chosen an implementation
in which the refrigerant ions are of a different atomic species,
with the relevant dipole transitions being separated by about
$10^5\:\mathrm{GHz}$ ($33\:\mathrm{nm}$).

The system we investigate is that of a two-ion crystal consisting
of one $\Be$ and one $\Mg$ ion.  The theoretical details of laser
cooling such a crystal can be found in \cite{cooling4,cooling3}.
The potential, in the small oscillation limit, is first expressed
in normal mode coordinates. In this representation the system is
that of six independent harmonic oscillators and, as such, laser
cooling a single normal mode using one ion is essentially the same
as cooling a single ion \cite{bible}. Here we restrict our
attention to the two modes involving axial motion only.  When the
masses are identical, these two modes are the center-of-mass (COM)
and stretch mode, in which the ions oscillate in phase and
$180^\circ$ out of phase respectively. When the masses are
different the two modes no longer correspond to center-of-mass and
relative motion. However, in what follows, we retain the terms COM
and stretch as the normal modes are still described by an in-phase
and out-of-phase motion analogous to the identical ion case
\cite{cooling3}.

\section{Sympathetic Cooling: Experimental Results}

The ions are confined in a linear Paul trap in which applied
static and RF potentials provide confinement \cite{Mary}. In the
typical setup used here the axial confinement results in COM and
stretch mode frequencies of $2.05\:\mathrm{MHz}$ and
$4.3\:\mathrm{MHz}$ respectively.  A quantization axis is
established with an applied static magnetic field of $B_0 \simeq
17\:\mathrm{G}$ which is oriented at an angle of $45^\circ$ with
respect to the trap axis. For completeness we have investigated
cooling of the two-ion crystal using either $\Be$ or $\Mg$ as the
refrigerant ion.

\subsection{Beryllium Cooling}

\begin{figure}
\includegraphics[height=4in]{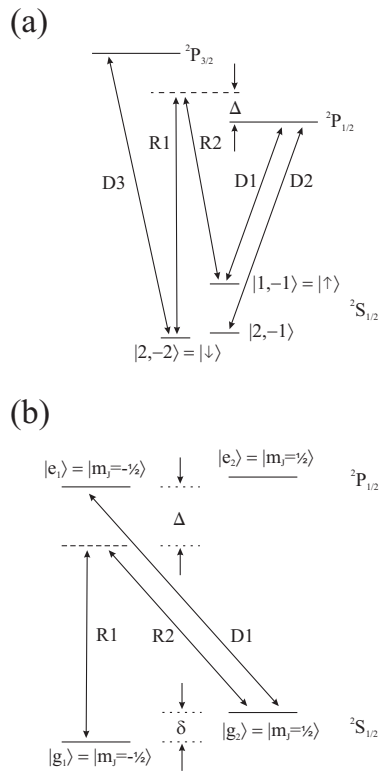}
\caption{\label{Levels} Relevant energy level structures for $\Be$
and $\Mg$ (not to scale) (a) Level structure for $\Be$. The
$^{2}S_{1/2}$ levels are labelled by their $F$, $m_F$ quantum
numbers.  Beams D1, D2, and D3 are all polarized $\sigma^{-}$.
Raman beams R1 and R2 are polarized $\pi$ and
$\sigma^{+}/\sigma^{-}$ respectively with a detuning $\Delta
\approx 2\pi \times 80\:\mathrm{GHz}$. For all beams
$\lambda\simeq 313\;\mathrm{nm}.$ (b) Level structure for $\Mg$.
Energy levels correspond to the $\ket{J=\frac{1}{2},m_J=\pm
\frac{1}{2}}$ states. D1 is near resonant and polarized
$\sigma^{-}$.  Raman beams R1 and R2 are polarized $\pi$ and
$\sigma^{+}/\sigma^{-}$ respectively, with a detuning $\Delta
\approx 2\pi \times 750\:\mathrm{MHz}$. The $17\:\mathrm{G}$ field
defining the quantization axis gives a ground state Zeeman
splitting of $\delta \approx 2\pi\times 40\:\mathrm{MHz}$.  For
all beams $\lambda \approx 280\:\mathrm{nm}$.}
\end{figure}

\begin{figure}
\includegraphics[height=4in]{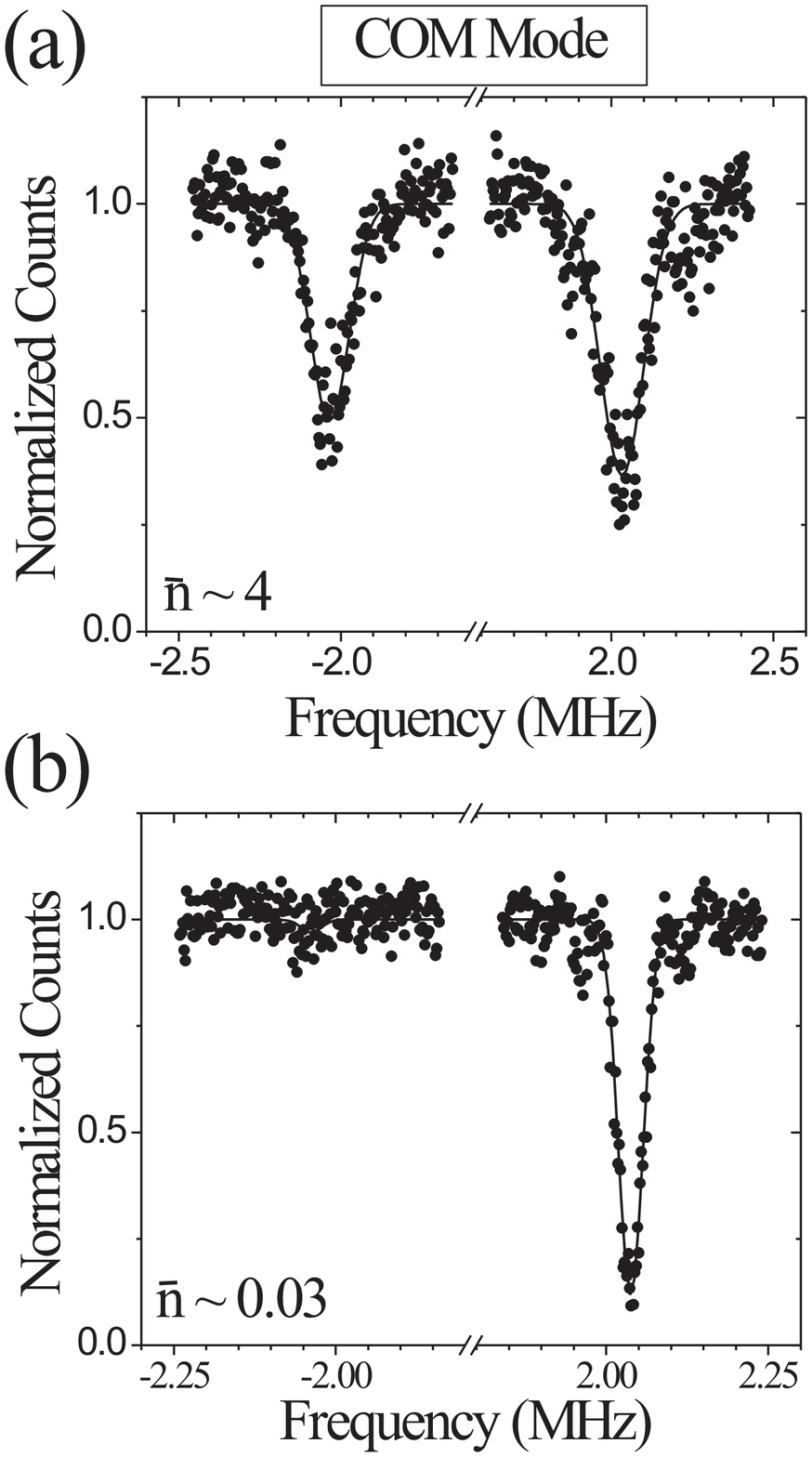}
\caption{\label{BeCOM} Results for the COM mode using $\Be$ as a
cooling ion. Red (blue) sideband data is shown on the left
(right). (a) Doppler cooling and (b) Sideband cooling. Values of
$\bar{n}$ are estimated from the sideband ratio - see text.}
\end{figure}

\begin{figure}
\includegraphics[height=4in]{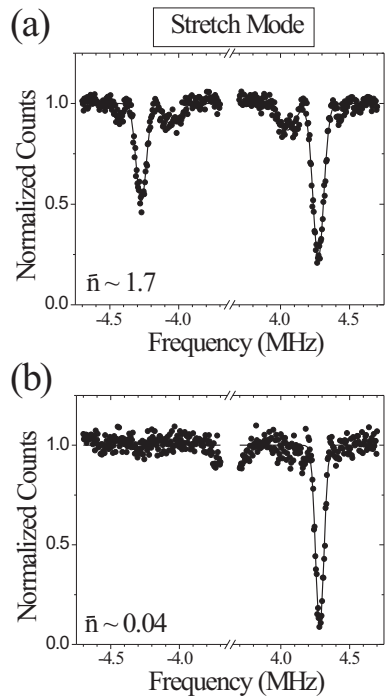}
\caption{\label{BeSTRETCH} Results for the stretch mode using
$\Be$ as a cooling ion.  Red (blue) sideband data is shown on the
left (right). (a) Doppler cooling and (b) Sideband cooling.
Values of $\bar{n}$ are estimated from the sideband ratio - see
text.}
\end{figure}

For $\Be$ the relevant level structure is shown in
Fig.~\ref{Levels}(a). Doppler cooling \cite{doppler, book2} is
achieved using the $\sigma^{-}$ polarized D3 beam, with beams D1
and D2 providing any necessary optical pumping (both of which are
also polarized $\sigma^{-}$).  Sideband cooling is then achieved
using beams R1 ($\pi$) and R2 ($\sigma^+/\sigma^-$) to drive
stimulated Raman transitions from $\ket{\downarrow,n}$ to
$\ket{\uparrow,n-1}$, followed by an optical pumping pulse
provided by beams D1 and D2 \cite{cooling5}. The Raman beams
propagate at right angles to each other with R2 parallel to the
quantization axis and the difference vector
$\Delta\mathbf{k}=\mathbf{k}_1-\mathbf{k}_2$ parallel to the trap
axis.  After 30 cooling cycles the population of the crystal's
motional ground state is probed using the $\Be$ ion as discussed
in \cite{cooling5,Heating}. Briefly, we drive Raman transitions on
the red and blue sideband for a fixed duration followed by a
detection pulse (beam D3), which measures the probability of being
in the $\ket{\downarrow}$ state. Assuming a thermal distribution
over the vibrational levels, the ratio, $r$, of the red and blue
sideband signal strengths yields a direct measure of the ground
state population and mean vibrational quanta via $P(n=0)=1-r$ and
$\bar{n}=r/(1-r)$ \cite{Heating}. In Figs.~\ref{BeCOM} and
\ref{BeSTRETCH} we show cooling results for both the stretch and
COM mode and for comparison we have included results from Doppler
cooling alone. From the sideband data we infer a ground state
occupancy of $0.97(2)$ for the COM mode and $0.96(3)$ for the
stretch with corresponding mean vibrational quanta of
$\bar{n}=0.03(2)$ and $\bar{n}=0.04(3)$ respectively.

\subsection{Magnesium Cooling}

\begin{figure}
\includegraphics[height=4in]{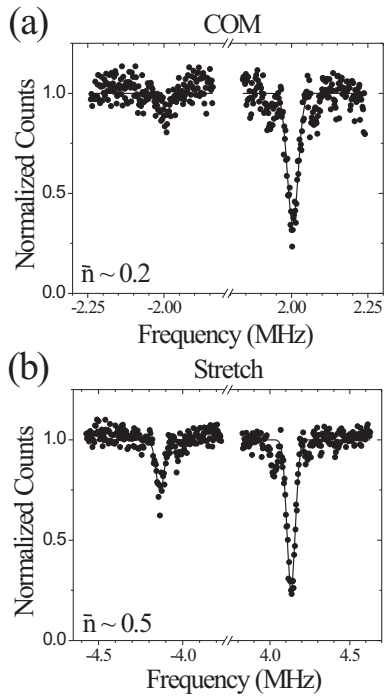}
\caption{\label{Mg} Cooling results using $\Mg$ as a cooling ion.
Red (blue) sideband data is shown on the left (right). (a) COM
mode and (b) Stretch mode. Values of $\bar{n}$ are estimated from
the sideband ratio - see text.}
\end{figure}

For $\Mg$ the relevant level structure is shown in
Fig.~\ref{Levels}(b). R2 is linearly polarized and propagates
along the quantization axis giving equal contributions to its
$\sigma^{+}/\sigma^{-}$ components.  This choice, together with
the $\pi$ polarized R1 beam, serves to partially balance the ac
Stark shifts induced by the Raman beams \footnote{There is a
residual differential stark shift due to the Zeeman splitting of
the excited and ground state manifolds.}. In addition, the
intensity of the R1 beam is chosen so that it has the same
coupling, $\Omega$ (Eq.~(\ref{eq:Hamiltonian})), to the excited
states as the $\sigma$ components of the R2 beam. This choice
gives the most favorable ratio of Raman coupling to spontaneous
emission rate for this beam configuration. While a linearly
polarized field tuned to the red of the
$^2S_{1/2}\leftrightarrow\,^2P_{1/2}$ transition could provide
Doppler cooling, this would require the use of an additional laser
beam. To circumvent this need the initial Doppler cooling step is
provided using $\Be$ as before.  For sideband cooling, Raman
transitions of a fixed duration are driven by beams R1 and R2,
which are tuned to the $\ket{g_1,n}\leftrightarrow\ket{g_2,n-1}$
transition and oriented similarly to the Raman beams used for
$\Be$. Each Raman pulse is then followed by an optical pumping
pulse provided by the near resonant $\sigma^{-}$ polarized beam
D1. Since the $^2S_{1/2} \leftrightarrow\,^2P_{1/2}$ manifold
lacks a closed cycling transition, we use $\Be$ as before in order
to probe the final ground state fraction and results are given in
Fig.~\ref{Mg}. From this data we infer a ground state occupancy of
$0.84(4)$ for the COM mode and $0.66(3)$ for the stretch mode,
with corresponding mean vibrational quanta of $\bar{n}=0.19(6)$
and $\bar{n}=0.52(7)$ respectively.

The significant difference between the results obtained using
$\Be$ as a refrigerant ion and those using $\Mg$ can be
predominately attributed to spontaneous emission from the $\Mg$
Raman cooling beams. For $\Be$, the Raman beams are detuned by
$\Delta \approx 2\pi\times 80\:\mathrm{GHz}$ and spontaneous
emission plays a negligibly small role. However, for $\Mg$, all
beams are derived from a single laser source with the Raman
detuning achieved using acousto-optic modulators. As such our
current setup limits the detuning of the Raman beams to
$\Delta\approx 2\pi\times 750\:\mathrm{MHz}$, and this gives rise
to a significant amount of spontaneous emission during a Raman
cooling cycle. Specifically, one can show that the mean number of
photons scattered in a time corresponding to a coherent $\pi$
pulse on the $\ket{g_1,n=1}\leftrightarrow \ket{g_2,n=0}$
transition is $0.55$ for the COM mode and $2.0$ for the stretch.
With spontaneous emission, population cannot be completely
transferred from one internal ground state to the other, and the
ground state $\ket{g_1,n=0}$ is no longer a dark state for the
Raman cooling process.  Thus, in any cooling cycle, population
cannot be completely transferred to the vibrational ground state.
Furthermore, after the Raman pulse, there will always be a finite
amount of population left in the internal state, $\ket{g_2}$. This
gives rise to a small amount of recoil heating during the final
optical pumping step.

\section{Simulations}

To more firmly establish spontaneous emission as the principal
limitation when cooling with $\Mg$ we have used a master equation
to model the cooling process, the details of which are given in
the appendices. In these simulations we have used
$\Omega=2\pi\times 30\:\mathrm{MHz}$, based on the measured beam
intensities, which yields a carrier Rabi frequency associated with
the Raman $\ket{g_1,n=0}\leftrightarrow\ket{g_2,n=0}$ transition
of approximately $2 \pi \times 300\:\mathrm{kHz}$.  Raman pulse
times of $2\:\mu\mathrm{s}$ and $5\:\mu\mathrm{s}$ were used, as
in the experiments, for the COM and stretch mode respectively. The
initial state was taken to be a thermal state in both cases with
$\bar{n}=4$ for the COM and $\bar{n}=1.7$ for the stretch,
consistent with the experimental data in Figs.~\ref{BeCOM}(a) and
~\ref{BeSTRETCH}(a).

\begin{figure}
\includegraphics[height=4in]{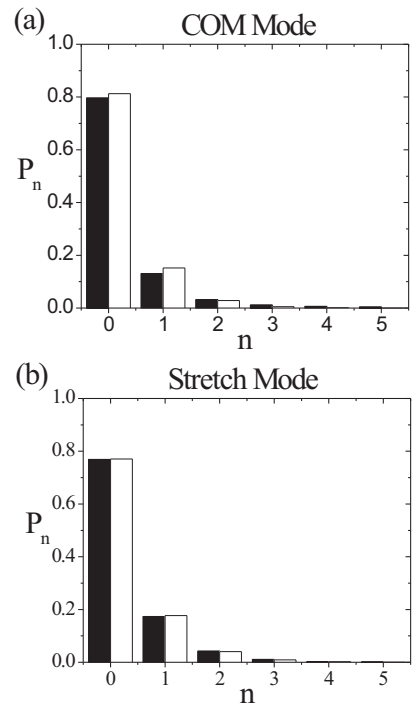}
\caption{\label{sims} Simulation results using the master equation
showing the final populations, $P_n$, vs. the respective
vibrational quanta, $n$. (a) COM mode and (b) Stretch mode. Black
bars correspond to the simulation results and the white bars are
the estimated thermal distributions based on sideband ratios.}
\end{figure}

The final distributions obtained from numerical integration of the
master equation are given in Fig.~\ref{sims}.   For these
distributions we find ground state occupancies of $0.80$ for the
COM mode and $0.77$ for the stretch mode with corresponding mean
vibrational quanta of $\bar{n}=0.46$ and $\bar{n}=0.31$
respectively.  These values of $\bar{n}$ are significantly
different from those found in the experiment.  However the
distributions in Fig.~\ref{sims} are not strictly thermal and are
not well characterized by $\bar{n}$.  Recall that we
experimentally characterize the final state based on the measured
sideband ratio and the assumption of a thermal state. Therefore,
to make a better comparison to the experiment, we calculate the
sideband ratios for the distributions in Fig.~\ref{sims} and we
find $r=0.18$ and $r=0.23$ for the COM and stretch respectively.
These values, together with the assumption of a thermal
distribution, give ground state occupancies of $0.82$ for the COM
mode and $0.77$ for the stretch with corresponding mean
vibrational quanta of $\bar{n}=0.23$ and $\bar{n}=0.30$
respectively.  These values are in better agreement with
experiment and, for comparison, we have included the estimated
thermal distributions in Fig.~\ref{sims}.

From the measured values of the frequencies and a normal mode
analysis we calculate Lamb-Dicke parameters of $0.3$ for the COM
mode and $0.082$ for the stretch. From these values, together with
the carrier frequency quoted above, one might expect pulse times
of $2.8\:\mu\mathrm{s}$ and $10\:\mu\mathrm{s}$ for the COM and
stretch respectively based on calculated $\pi$ times for the
$\ket{g_1,n=1}\leftrightarrow \ket{g_2,n=0}$ transition. The
significant difference between these pulse times and those used in
both the simulations and the experiments is due to that fact that
we do not achieve ground state cooling.  Thus the final optimum
pulse need not be a $\pi$ pulse on $\ket{g_1,n=1}\leftrightarrow
\ket{g_2,n=0}$ transition.

\section{RF Micromotion Compensation}

An additional technical advantage of using different atomic
species arises from the difference in mass of the two ions. In a
linear Paul trap \cite{trap,bible} the transversal confinement
provided by the applied RF fields can be described by a harmonic
pseudo-potential with a frequency that is inversely proportional
to the mass of the ion. Thus, in the presence of a transverse
stray electric field, the two ions have an equilibrium position
that depends on the mass and their relative position vector no
longer lies parallel to the trap axis.  The net effect is that the
modes are no longer separable into the axial and radial
directions, and the measured frequencies deviate from those found
in the absence of stray fields. Since this effect only occurs in
the presence of stray fields, it provides an experimental method
to detect and eliminate the RF-micromotion they induce.

To determine the effectiveness of this method we compare it with a
previously demonstrated method in which the micromotion is
monitored by measuring the below saturation scattering rate $R_0$
on the cycling transition in $\Be$ (D3 in Fig.~\ref{Levels}(a))
when the laser is tuned near the carrier
($\omega_{\mathrm{laser}}-\omega_{\mathrm{atom}}\simeq 0$) and the
scattering rate $R_1$ when the laser is tuned near the first
micromotion-induced sideband
($\omega_{\mathrm{laser}}-\omega_{\mathrm{atom}}\simeq \pm
\Omega_{RF}$) \cite{Micromotion}.  From these measurements the
amplitude of the micromotion can be found from the ratio
\begin{eqnarray}
\label{eq:fluorescence}
R=\frac{R_1}{R_0}=\frac{J_1^2(\mathbf{k}\cdot\mathbf{u})}{J_0^2(\mathbf{k}\cdot\mathbf{u})}
\end{eqnarray}
where $J_k$ are Bessel functions of the first kind, $\mathbf{k}$
is the wave vector of the laser and $\mathbf{u}$ is the
micromotion amplitude.  To compare the two techniques we determine
the effect on the frequency spectrum for a given $R$.

In the pseudo-potential approximation, the potential energy for a
$\Be$ ion of mass $m$ can be written
\begin{eqnarray}
\label{Bepot} V=\frac{m}{2}\bigg[\omega_1^2
x_1^2+\left(\omega_0^2+\omega_2^2 \right)
x_2^2+\left(\omega_0^2-\omega_3^2 \right) x_3^2\bigg],
\end{eqnarray}
where $\omega_0$ is the frequency associated with the RF
pseudo-potential and $\omega_1^2$, $\omega_2^2$, and $-\omega_3^2$
are the curvatures associated with the applied static field needed
to provide confinement along the trap axis $(\mathbf{\hat{x}}_1)$
\footnote{The sign of the static field curvatures depends on how
the field is applied.  For our parameters the two radial
curvatures have opposite sign as indicated in Eq~(\ref{Bepot}).}.
Since $\omega_0$ and the static field curvatures $\omega_k^2$ are
inversely proportional to the mass the potential energy of the two
ion system can be written
\begin{eqnarray}
\label{eq:potential}
V&=&\frac{m}{2}\bigg[\omega_1^2
x_1^2+\left(\omega_0^2+\omega_2^2 \right)
x_2^2+\left(\omega_0^2-\omega_3^2 \right)
x_3^2\bigg]\nonumber\\
& & +\frac{m}{2}\bigg[\omega_1^2 y_1^2+\left(\mu^{-1}
\omega_0^2+\omega_2^2 \right)
y_2^2\nonumber\\
& &+\left(\mu^{-1}\omega_0^2-\omega_3^2 \right)
y_3^2\bigg]+\frac{q^2}{4\pi
\epsilon_0}\frac{1}{|\mathbf{x}-\mathbf{y}|}\nonumber\\
& &+a m
\omega_0^2\left[(x_2+y_2)\cos{\theta}+(x_3+y_3)\sin{\theta}\right],
\end{eqnarray}
where $\mathbf{x}$ and $\mathbf{y}$ denote the positions of $\Be$
and $\Mg$ respectively, $\mu$ is the ratio of the $\Mg$ mass to
the $\Be$ mass, and $q$ is the ion's charge.  In this expression
we have included a stray electric field
\begin{eqnarray}
\label{stray} \mathbf{E}=a\frac{m\,\omega_0^2}{q}(\cos{\theta}
\:\mathbf{\hat{x}}_2+\sin{\theta}\:\mathbf{\hat{x}}_3).
\end{eqnarray}
In the limit that $\omega_0$ dominates the radial confinement, $a$
is the off-axis displacement of $\Be$ due to the stray field.
More generally it can be shown \cite{Micromotion} that an electric
field of this form leads to micromotion of amplitude
\begin{eqnarray}
\mathbf{u}=a \frac{\omega_0
\sqrt{2}}{\Omega_{RF}}\left(\frac{\omega_0^2}{\omega_0^2+\omega_2^2}
\cos{\theta}\:\mathbf{\hat{x}}_2+\frac{\omega_0^2}{\omega_0^2-\omega_3^2}
\sin{\theta}\:\mathbf{\hat{x}}_3\right).
\end{eqnarray}
For our particular setup we have $\omega_0 \simeq 2\pi\times
9\:\mathrm{MHz}$, $\omega_1 \simeq 2\pi\times 2.8\:\mathrm{MHz}$,
$\omega_2 \simeq 2\pi\times 2\:\mathrm{MHz}$, and $\omega_3 \simeq
2\pi\times 3.4\:\mathrm{MHz}$ and so we neglect the small
dependence of the micromotion amplitude on the applied static
field.  If we assume a probe beam is optimally aligned for
micromotion detection we have
\begin{eqnarray}
\label{eq:micromotion}
\mathbf{k}\cdot\mathbf{u}=\frac{2\pi}{\lambda}\frac{\omega_0
\sqrt{2}}{\Omega_{RF}} a.
\end{eqnarray}

\begin{figure}
\includegraphics[height=4in]{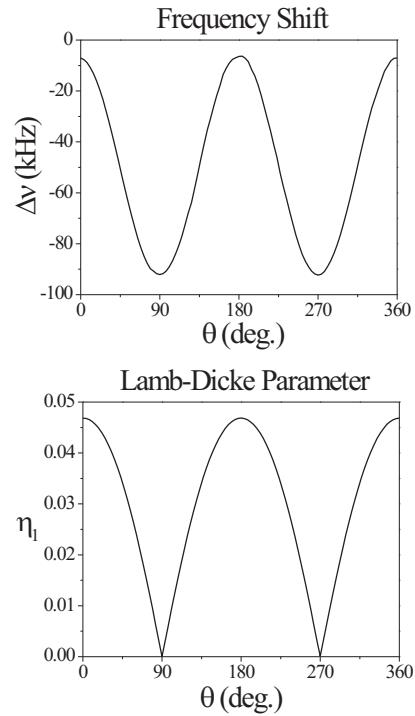}
\caption{\label{fshift} Plots showing effects of a transverse
stray field on frequency spectrum for our experimental parameters.
(a) Stretch mode shift and (b) Lamb-Dicke parameter for radial
rocking mode in the $\mathbf{\hat{x}}_2$ as a function of the
angle $\theta$.}
\end{figure}

Using this expression in Eq.~(\ref{eq:fluorescence}) we can then
find $a$ for a given ratio $R$.  Using this value of $a$ in
Eq.~(\ref{eq:potential}) we can find, for a given $\theta$, the
equilibrium position and normal mode frequencies for the two ion
system \cite{Goldstein}.  In Fig.~\ref{fshift}(a) we plot the
shift in the stretch mode frequency as a function of $\theta$
using $R=0.1$ and the experimental value
$\Omega_{RF}=2\pi\times110\:\mathrm{MHz}$.  The maximum frequency
shift of $90\:\mathrm{kHz}$ is easily detected from the difference
in the stretch mode sideband and carrier resonant frequencies. The
minimum frequency shift of $7\:\mathrm{kHz}$ on the other hand
would indicate a reduced sensitivity to stray fields acting along
the $\hat{x}_2$ direction.  The sensitivity could be increased by
applying a static potential with negative curvature in this
direction.  This could be accomplished by, for example, applying
an overall static potential difference between the RF and (RF
grounded) control electrodes \cite{Mary}.  Alternatively we could
utilize an additional effect that influences the observed Raman
frequency spectrum. The Raman beams used to probe the mode
frequencies couple to all modes that have motion along the
direction of the difference vector $\Delta
\mathbf{k}=\mathbf{k}_1-\mathbf{k}_2$ which, in this case, is
aligned along the trap axis.  Thus, in the presence of a stray
field, the Raman beams can couple to other modes giving rise to
additional features in the frequency spectrum.  The strength of
this coupling is determined by the Lamb-Dicke parameter, $\eta_1$,
and in Fig.~\ref{fshift}(b) we plot the magnitude of $\eta_1$ as a
function of $\theta$ for the radial rocking mode whose motion is
nominally in the $\mathbf{\hat{x}}_2$ direction.  The maximum
value of $\eta_1 \simeq 0.05$ gives rise to an easily detected
feature in the frequency spectrum.  Since this effect complements
the shift of the stretch mode frequency, compensation of stray
field from any direction can be accomplished.

Although the sensitivity of this approach depends on the strength
of the RF confinement, it does have advantages over fluorescence
measurements where, to null micromotion in the transverse
direction, we require two non-parallel probe laser beam
directions. In particular, monitoring the motional frequency
spectrum does not require any additional optical access and is
sensitive to micromotion in all transverse directions.
Furthermore, the mode frequencies are needed for quantum logic
experiments anyway and so monitoring the spectrum is a natural
part of the experiment setup. As such, we have found this property
to be a convenient method to detect and compensate the presence of
stray fields.

\section{Spontaneous Emission and Stark Shifts}

The advantage of using a different atomic species for the
refrigerant ion is in the small Stark shifts and off-resonant
excitation of the qubit ion due to the presence of the cooling
light.  Although the spontaneous emission and Stark shifts on the
qubit ion will depend on the exact experimental conditions, we can
get an estimate of these effects with the following simple model.
We calculate the probability of spontaneous emission from a qubit
based on a superposition of the $|\!\downarrow \rangle$ and
$|\!\uparrow \rangle$ states of $\Be$, during a $\pi$ pulse on the
$|g_1, n = 1\rangle \leftrightarrow |g_2, n=0 \rangle$ $\Mg$ Raman
cooling transition. (This will be the dominant source of
spontaneous emission during one cooling cycle since the re-pumping
pulse is assumed to be near resonant and therefore of much less
intensity.) We assume the laser beam intensity is the same on both
ions. The time for the cooling pulse is given by $\tau_{\pi}
\simeq 2 \pi \Delta /(\eta \Omega^2)$. From Ref. \cite{Wineland},
we find the rate of spontaneous emission from the $^9$Be$^+$ qubit
to be $R_{SE} = (3/2) \gamma \Omega^2 /(\Delta^*)^2$ where
$\Delta^*$ is the frequency difference between the relevant
transitions for $^9$Be$^+$ and $^{24}$Mg$^+$. Therefore, the
probability of spontaneous emission from the qubit superposition
is given by $P_{SE} = R_{SE}\tau_{\pi} = 3\pi \gamma \Delta/(\eta
(\Delta^*)^2)$. For the COM mode ($\eta = 0.3$), using the
experimental values from Sec. III, we find $P_{SE} \simeq
4\times10^{-11}$. Better cooling with $\Mg$ does require us to
increase the Raman beam detuning, which leads to a corresponding
increase in the probability of spontaneous emission (proportional
to $\Delta$), but this should still yield a negligible probability
of spontaneous emission.

To estimate Stark shifts on the qubit, we first note that by using
linearly polarized light to drive the $\Mg$ Raman cooling
transition, net Stark shifts on the $\Be$ qubit will vanish
\cite{Wineland}. However, during the $\Mg$ re-pumping pulse, we
require $\sigma^+$ light, which leads to Stark shifts on the
qubit. From Ref. \cite{Wineland}, we estimate the phase shift on
the qubit for each re-pumping pulse to be approximately
$\Delta\phi \simeq 3\times10^{-10}$.  Because the expected
decoherence from spontaneous emission and Stark shifts are so
small, we could not test these predictions.  Nevertheless, we did
look for decoherence due to presence of the $\Mg$ Raman cooling
beams.  We performed a Ramsey resonance experiment on the
$|\downarrow\rangle \rightarrow |\uparrow\rangle$ transition in
$\Be$ with a precession time of 10 ms and observed no loss in
contrast ($5\%$ accuracy) with the $\Mg$ Raman beams applied
continuously with maximum intensity during this time.

\section{Conclusion}

We have demonstrated sympathetic cooling of a two ion crystal
consisting of one $\Be$ and one $\Mg$. Using $\Be$ as the cooling
ion, we have been able to achieve a ground state occupancy for the
axial modes of $\sim 95\%$ and better cooling results can be
expected with improvements in the apparatus. With $\Mg$ as the
cooling ion we are currently limited by spontaneous emission. A
model has been developed which takes spontaneous emission into
account and gives reasonable agreement with the experimental
results. Thus, for $\Mg$ cooling, we expect to find a substantial
improvement in the cooling results when we use larger detunings.
For the purposes of quantum information processing one can
anticipate the need to cool more than two ions: the three ion
crystal consisting of two qubits and one refrigerant being an
obvious example.  For such cases the salient features of the
cooling process are essentially the same as those demonstrated
here, provided the normal mode frequencies can be spectrally
resolved.  This spectral resolution will typically be achievable
in the case where two adjacent qubit ions are cooled by a third
refrigerant ion \cite{bible,Kielpinski}.  A configuration where an
even number of qubit ions is cooled by a centrally located
refrigerant ion \cite{Kielpinski} should also be experimentally
feasible. Finally, we note that these techniques may find use
outside the realm of quantum computation, for instance, when
applied to atomic clocks \cite{added3}.

In addition to demonstrating sympathetic cooling, we have found
that a mass difference between ions gives rise to a novel
technique to detect and eliminate the effects of RF micromotion.
This technique involves monitoring the motional frequency spectrum
only and, as such, it is a very convenient and easily implemented
method of micromotion compensation.

The authors thank Marie Jensen and Piet O. Schmidt for suggestions
and comments on the manuscript.  This work was supported by the
U.S. National Security Agency (NSA) and Advanced Research and
Development Activity (ARDA) under contract No. MOD-7171.00, the
U.S. Office of Naval Research (ONR), and the National Institute of
Standards and Technology (NIST), an agency of the U.S. government.
This paper is a contribution of NIST and is not subject to U.S.
copyright.

\appendix

\section{\label{appA}Model}
In the interaction picture the master equation has the usual form
\begin{eqnarray}
\label{eq:master}
\dot{\rho}=-i\left[H_I,\rho\right]+L\rho
\end{eqnarray}
where $H_I$ is the interaction Hamiltonian and $L$ is the
Liouvillian operator which accounts for dissipative processes. For
the $\Mg$ system depicted in Fig.~\ref{Levels}(b) the Hamiltonian
is given by
\begin{eqnarray}
\label{eq:Hamiltonian}
H_{I}&=& \frac{\Omega}{2}\,e^{i \Delta
t}\bigg[D\left(e^{i\delta_L t} \ket{e_1}\bra{g_1} +
e^{i(\delta_L-\frac{2}{3}\delta)t} \ket{e_2}
\bra{g_2}\right)\nonumber\\
& &+D^{\dag}\left( \ket{e_1}\bra{g_2}+e^{i\frac{4}{3}\delta
t}\ket{e_2}\bra{g_1}\right)\bigg]+c.c.
\end{eqnarray}
where $\delta_L=\omega_{R2}-\omega_{R1}-\delta$ is the relative
detuning of the Raman beams from the Raman resonance and $D$ is
the kick or recoil operator. In terms of the annihilation
operator, $a$, the normal mode frequency, $\omega$, and the
corresponding Lamb-Dicke parameter, $\eta$, this operator may be
written
\begin{eqnarray}
D=\exp\left[i \frac{\eta}{2}\left(a e^{-i\omega t}+a^{\dag}e^{i
\omega t}\right)\right].
\end{eqnarray}
where we have restricted our attention to the mode of interest and
$\eta$ is defined in terms of $\Delta \mathbf{k}$ for the Raman
beams. Using the notation $\sigma_{jk}=\ket{g_j}\bra{e_k}$ the
Liouvillian has the usual form
\begin{eqnarray}
\label{eq:Liouvillian} L\rho=\frac{\Gamma}{2} \sum\limits_{j,k}
c_{jk}^{2} (2\sigma_{jk}\bar{\rho}\sigma_{jk}^{\dag} -
\sigma_{jk}^{\dag}\sigma_{jk} \rho-\rho
\sigma_{jk}^{\dag}\sigma_{jk})
\end{eqnarray}
where $c_{jk}$ is the Clebsch-Gordon coefficient connecting
$\ket{g_j}$ and $\ket{e_k}$,  $\Gamma$ is the total decay rate
from the excited state, and $\bar{\rho}$ describes the density
matrix after a spontaneous emission event \cite{recoil1,recoil2}:
\begin{eqnarray}
\bar{\rho}&=&\iint d\, \Omega \,W(\hat{k}\cdot\hat{e}_z)
\exp\left(i \eta\frac{\hat{k}\cdot\hat{e}_t}{\sqrt{2}}\left(a
e^{-i\omega t}+a^{\dag}e^{i \omega
t}\right)\right) \nonumber \\
& & \times \rho \exp\left(-i
\eta\frac{\hat{k}\cdot\hat{e}_t}{\sqrt{2}} \left(a e^{-i\omega
t}+a^{+}e^{i \omega t}\right)\right).
\end{eqnarray}
In this expression $\hat{k}$ is the unit vector giving the
propagation direction of the scattered photon relative to the
quantization axis denoted $\hat{e}_z$, $\hat{e}_t$ is a unit
vector along the trap axis, $W(\hat{k}\cdot\hat{e}_z)$ is the
angular distribution of the emission, and the integration is
carried over the full solid angle. For the dipole decay discussed
here we have
\begin{equation}
W(\hat{k}\cdot\hat{e}_z)=\frac{3}{8 \pi} \begin{cases}
\left(1-(\hat{k}\cdot\hat{e}_z)^2\right),\\
\frac{1}{2}\left(1+(\hat{k}\cdot\hat{e}_z)^2\right)
\end{cases}
\end{equation}
where the upper (lower) expression applies to decay channels
involving linear (circular) polarization.

\vspace{7mm}
\section{\label{appB}Adiabatic Elimination}

The master equation formulated in appendix \ref{appA} can be
integrated directly using a truncated basis of Fock states to
describe the vibrational modes. However it is computationally
intensive to do so and significant simplification can be achieved
by adiabatically eliminating the excited state. The procedure we
adopt here closely follows that found in \cite{book1}.

Let $P_g$ and $P_e$ be the projection operators defined by
$P_g=\sum \limits_{k} \ket{g_k}\bra{g_k}$ and $P_e=\sum
\limits_{k} \ket{e_k}\bra{e_k}$.  Then any operator, $A$, can be
written as $A=A_{ee}+A_{eg}+A_{ge}+A_{gg}$ where $A_{ab}=P_a A
P_b$.  In this way the master Eq.~(\ref{eq:master}) can be
rewritten in component form giving
\begin{eqnarray}
\label{eq:ae1}
\dot{\rho}_{ee}&=&-i
\left(\tilde{H}_{eg}\tilde{\rho}_{ge}-\tilde{\rho}_{eg}\tilde{H}_{ge}\right)-\Gamma
\rho_{ee} \\
\label{eq:ae2}
\dot{\tilde{\rho}}_{eg}&=&-i\left(\tilde{H}_{eg}\rho_{gg}-\rho_{ee}\tilde{H}_{eg}\right)
- \left(\frac{\Gamma}{2}+i\Delta\right)\tilde{\rho}_{eg} \\
\label{eq:ae3}
\dot{\rho}_{gg}&=&-i\left(\tilde{H}_{ge}\tilde{\rho}_{eg}-\tilde{\rho}_{ge}\tilde{H}_{eg}\right)
\nonumber \\
& & \qquad + \Gamma \sum \limits_{j,k} c_{jk}^{2} \sigma_{jk}
\bar{\rho}_{ee} \sigma_{jk}^{+},
\end{eqnarray}
where have used the definitions $H_{I}=e^{i \Delta
t}\tilde{H}_{eg}+c.c$ and $\tilde{\rho}_{eg}=e^{-i \Delta
t}\rho_{eg}.$  To proceed we neglect the the terms
$\dot{\tilde{\rho}}_{eg}$ and $\rho_{ee}\tilde{H}_{eg}$ in
Eq.~(\ref{eq:ae2}), the validity of which is discussed in
\cite{book1}. This results in an algebraic expression for
$\tilde{\rho}_{eg}$ which can be rearranged to give
\begin{eqnarray}
\label{eq:coherence} \tilde{\rho}_{eg}=\frac{-i}{(\Gamma/2)+i
\Delta} \tilde{H}_{eg} \rho_{gg}.
\end{eqnarray}
Substituting this expression into Eq.~(\ref{eq:ae1}) and
neglecting $\dot{\rho}_{ee}$ then yields the result
\begin{eqnarray}
\label{eq:excited} \rho_{ee}=\frac{1}{(\Gamma/2)^2+\Delta^2}
\tilde{H}_{eg}\rho_{gg}\tilde{H}_{ge}.
\end{eqnarray}
Finally, Eqs.~(\ref{eq:coherence}) and (\ref{eq:excited}) can be
substituted into Eq.~(\ref{eq:ae3}) to give a closed equation of
motion for $\rho_{gg}$.  Dropping the subscripts on $\rho_{gg}$ we
find the equation
\begin{widetext}
\begin{eqnarray}
\label{eq:master2} \dot{\rho}&=&
-i\left[\frac{-\Delta}{(\Gamma/2)^2+\Delta^2} \tilde{H}_{ge}
\tilde{H}_{eg},\rho\right]
-\left\{\frac{(\Gamma/2)}{(\Gamma/2)^2+\Delta^2} \tilde{H}_{ge}
\tilde{H}_{eg},\rho\right\} +
\frac{(\Gamma/2)}{(\Gamma/2)^2+\Delta^2} \sum \limits_{j,k}
c_{jk}^{2} 2 \sigma_{jk} \tilde{H}_{eg} \bar{\rho} \tilde{H}_{ge}
\sigma_{jk}^{\dag}
\end{eqnarray}
\end{widetext}
for the effective ground state density matrix, where $\{\:,\}$ is
used to denote an anti-commutator. If one makes the
identifications
\begin{align*}
H_{\mathrm{\textit{eff}}}&=\frac{-\Delta}{(\Gamma/2)^2+\Delta^2}
\tilde{H}_{ge} \tilde{H}_{eg},\\
\Gamma'&=\frac{\Gamma}{(\Gamma/2)^2+\Delta^2},\\
\intertext{and} \bar{\sigma}_{jk}&=\sigma_{jk} \tilde{H}_{eg}
\end{align*}
then Eq.~(\ref{eq:master2}) can be recast into the usual form
given in Eqs.~(\ref{eq:master}) and (\ref{eq:Liouvillian}).
Numerical simulation of the reduced master equation can be further
simplified by expanding out the terms in (\ref{eq:master2}) and
neglecting the off-resonant terms associated with the Raman pair
$\sigma^{+}$ and $\pi$.

The resulting system gives excellent agreement with the full
system given in Eq.~(\ref{eq:master}).  However the procedure
outlined here is not rigorously correct due the residual
time-dependence in $\tilde{H}_{eg}$ associated with the Zeeman
splitting of the excited and ground state manifolds. In effect,
the procedure here neglects the small differential Stark shift
induced by this splitting as well as the small change in the
various spontaneous emission rates.  These have been accounted for
in a more elaborate treatment in which the projection operators
are split allowing the individual level shifts to be accounted
for. The resulting algebra is more involved but the procedure is
precisely the same. For brevity and clarity we have omitted the
details of this more elaborate treatment.  Finally, we note that
the optical pumping process can be treated in a similar manner
with Eq.~(\ref{eq:Hamiltonian}) replaced with
\begin{eqnarray}
H_{I}=\frac{\Omega}{2} D^{\dag} \ket{e_1}\bra{g_2}+c.c.
\end{eqnarray}
\bibliography{Cooling}

\begin{thebibliography}{31}
\expandafter\ifx\csname natexlab\endcsname\relax\def\natexlab#1{#1}\fi
\expandafter\ifx\csname bibnamefont\endcsname\relax
  \def\bibnamefont#1{#1}\fi
\expandafter\ifx\csname bibfnamefont\endcsname\relax
  \def\bibfnamefont#1{#1}\fi
\expandafter\ifx\csname citenamefont\endcsname\relax
  \def\citenamefont#1{#1}\fi
\expandafter\ifx\csname url\endcsname\relax
  \def\url#1{\texttt{#1}}\fi
\expandafter\ifx\csname urlprefix\endcsname\relax\def\urlprefix{URL }\fi
\providecommand{\bibinfo}[2]{#2}
\providecommand{\eprint}[2][]{\url{#2}}

\bibitem[{\citenamefont{Cirac and Zoller}(1995)}]{gate1}
\bibinfo{author}{\bibfnamefont{J.~I.} \bibnamefont{Cirac}} \bibnamefont{and}
  \bibinfo{author}{\bibfnamefont{P.}~\bibnamefont{Zoller}},
  \bibinfo{journal}{Phys. Rev. Lett.} \textbf{\bibinfo{volume}{74}},
  \bibinfo{pages}{4091} (\bibinfo{year}{1995}).

\bibitem[{\citenamefont{Schmidt-Kaler et~al.}(2003)}]{added2}
\bibinfo{author}{\bibfnamefont{F.}~\bibnamefont{Schmidt-Kaler}}
  \bibnamefont{et~al.}, \bibinfo{journal}{Nature}
  \textbf{\bibinfo{volume}{422}}, \bibinfo{pages}{408} (\bibinfo{year}{2003}).

\bibitem[{\citenamefont{S{\o}rensen and M{\o}lmer}(1999)}]{gate2}
\bibinfo{author}{\bibfnamefont{A.}~\bibnamefont{S{\o}rensen}} \bibnamefont{and}
  \bibinfo{author}{\bibfnamefont{K.}~\bibnamefont{M{\o}lmer}},
  \bibinfo{journal}{Phys. Rev. Lett.} \textbf{\bibinfo{volume}{82}},
  \bibinfo{pages}{1971} (\bibinfo{year}{1999}).

\bibitem[{\citenamefont{S{\o}rensen and M{\o}lmer}(2000)}]{gate3}
\bibinfo{author}{\bibfnamefont{A.}~\bibnamefont{S{\o}rensen}} \bibnamefont{and}
  \bibinfo{author}{\bibfnamefont{K.}~\bibnamefont{M{\o}lmer}},
  \bibinfo{journal}{Phys. Rev. A} \textbf{\bibinfo{volume}{62}},
  \bibinfo{pages}{022311} (\bibinfo{year}{2000}).

\bibitem[{\citenamefont{Milburn et~al.}(2000)\citenamefont{Milburn, Schneider,
  and James}}]{gate5}
\bibinfo{author}{\bibfnamefont{G.~J.} \bibnamefont{Milburn}},
  \bibinfo{author}{\bibfnamefont{S.}~\bibnamefont{Schneider}},
  \bibnamefont{and} \bibinfo{author}{\bibfnamefont{D.~F.~V.}
  \bibnamefont{James}}, \bibinfo{journal}{Fortschritte der Physik}
  \textbf{\bibinfo{volume}{48}}, \bibinfo{pages}{801} (\bibinfo{year}{2000}).

\bibitem[{\citenamefont{Cirac and Zoller}(2000)}]{gate6}
\bibinfo{author}{\bibfnamefont{J.~I.} \bibnamefont{Cirac}} \bibnamefont{and}
  \bibinfo{author}{\bibfnamefont{P.}~\bibnamefont{Zoller}},
  \bibinfo{journal}{Nature} \textbf{\bibinfo{volume}{404}},
  \bibinfo{pages}{579} (\bibinfo{year}{2000}).

\bibitem[{\citenamefont{Calaraco et~al.}(2001)\citenamefont{Calaraco, Cirac,
  and Zoller}}]{gate7}
\bibinfo{author}{\bibfnamefont{T.}~\bibnamefont{Calaraco}},
  \bibinfo{author}{\bibfnamefont{J.~I.} \bibnamefont{Cirac}}, \bibnamefont{and}
  \bibinfo{author}{\bibfnamefont{P.}~\bibnamefont{Zoller}},
  \bibinfo{journal}{Phys. Rev. A} \textbf{\bibinfo{volume}{63}},
  \bibinfo{pages}{062304} (\bibinfo{year}{2001}).

\bibitem[{\citenamefont{James}(2000)}]{gate8}
\bibinfo{author}{\bibfnamefont{D.~F.~V.} \bibnamefont{James}}, in
  \emph{\bibinfo{booktitle}{Scalable Quantum Computers}}, edited by
  \bibinfo{editor}{\bibfnamefont{S.~L.} \bibnamefont{Braunstein}},
  \bibinfo{editor}{\bibfnamefont{H.~K.} \bibnamefont{Lo}}, \bibnamefont{and}
  \bibinfo{editor}{\bibfnamefont{P.}~\bibnamefont{Kok}}
  (\bibinfo{publisher}{Wiley-VCH}, \bibinfo{address}{Berlin},
  \bibinfo{year}{2000}), pp. \bibinfo{pages}{53--68}.

\bibitem[{\citenamefont{Sackett et~al.}(2000)}]{gate9}
\bibinfo{author}{\bibfnamefont{C.~A.} \bibnamefont{Sackett}}
  \bibnamefont{et~al.}, \bibinfo{journal}{Nature}
  \textbf{\bibinfo{volume}{404}}, \bibinfo{pages}{256} (\bibinfo{year}{2000}).

\bibitem[{\citenamefont{Leibfried et~al.}(2003)}]{gate4}
\bibinfo{author}{\bibfnamefont{D.}~\bibnamefont{Leibfried}}
  \bibnamefont{et~al.}, \bibinfo{journal}{Nature}
  \textbf{\bibinfo{volume}{422}}, \bibinfo{pages}{412} (\bibinfo{year}{2003}).

\bibitem[{\citenamefont{Wineland et~al.}(1998)}]{bible}
\bibinfo{author}{\bibfnamefont{D.~J.} \bibnamefont{Wineland}}
  \bibnamefont{et~al.}, \bibinfo{journal}{J. Res. Natl. Inst. Stand. Technol.}
  \textbf{\bibinfo{volume}{103}}, \bibinfo{pages}{259} (\bibinfo{year}{1998}).

\bibitem[{\citenamefont{Kielpinski et~al.}(2002)\citenamefont{Kielpinski,
  Monroe, and Wineland}}]{Kielpinski}
\bibinfo{author}{\bibfnamefont{D.}~\bibnamefont{Kielpinski}},
  \bibinfo{author}{\bibfnamefont{C.}~\bibnamefont{Monroe}}, \bibnamefont{and}
  \bibinfo{author}{\bibfnamefont{D.~J.} \bibnamefont{Wineland}},
  \bibinfo{journal}{Nature} \textbf{\bibinfo{volume}{417}},
  \bibinfo{pages}{709} (\bibinfo{year}{2002}).

\bibitem[{\citenamefont{Rowe et~al.}(2002)}]{Mary}
\bibinfo{author}{\bibfnamefont{M.~A.} \bibnamefont{Rowe}} \bibnamefont{et~al.},
  \bibinfo{journal}{Quant. Inf. and Comp.} \textbf{\bibinfo{volume}{2}},
  \bibinfo{pages}{257} (\bibinfo{year}{2002}).

\bibitem[{\citenamefont{Larson et~al.}(1986)}]{added1}
\bibinfo{author}{\bibfnamefont{D.~J.} \bibnamefont{Larson}}
  \bibnamefont{et~al.}, \bibinfo{journal}{Phys. Rev. Lett.}
  \textbf{\bibinfo{volume}{57}}, \bibinfo{pages}{70} (\bibinfo{year}{1986}).

\bibitem[{\citenamefont{Rohde et~al.}(2001)}]{cooling1}
\bibinfo{author}{\bibfnamefont{H.}~\bibnamefont{Rohde}} \bibnamefont{et~al.},
  \bibinfo{journal}{J. Opt. B.} \textbf{\bibinfo{volume}{3}},
  \bibinfo{pages}{S34} (\bibinfo{year}{2001}).

\bibitem[{\citenamefont{Steane et~al.}(2000)}]{Steane}
\bibinfo{author}{\bibfnamefont{A.}~\bibnamefont{Steane}} \bibnamefont{et~al.},
  \bibinfo{journal}{Phys. Rev. A} \textbf{\bibinfo{volume}{62}},
  \bibinfo{pages}{042305} (\bibinfo{year}{2000}).

\bibitem[{\citenamefont{Blinov et~al.}(2002)\citenamefont{Blinov, Deslauriers,
  Lee, Madsen, Miller, and Monroe}}]{cooling2}
\bibinfo{author}{\bibfnamefont{B.~B.} \bibnamefont{Blinov}},
  \bibinfo{author}{\bibfnamefont{L.}~\bibnamefont{Deslauriers}},
  \bibinfo{author}{\bibfnamefont{P.}~\bibnamefont{Lee}},
  \bibinfo{author}{\bibfnamefont{M.~J.} \bibnamefont{Madsen}},
  \bibinfo{author}{\bibfnamefont{R.}~\bibnamefont{Miller}}, \bibnamefont{and}
  \bibinfo{author}{\bibfnamefont{C.}~\bibnamefont{Monroe}},
  \bibinfo{journal}{Phys.\ Rev. A} \textbf{\bibinfo{volume}{65}},
  \bibinfo{pages}{040304} (\bibinfo{year}{2002}).

\bibitem[{\citenamefont{Kelpinski et~al.}(2000)}]{cooling4}
\bibinfo{author}{\bibfnamefont{D.}~\bibnamefont{Kelpinski}}
  \bibnamefont{et~al.}, \bibinfo{journal}{Phys.\ Rev. A}
  \textbf{\bibinfo{volume}{61}}, \bibinfo{pages}{032310}
  (\bibinfo{year}{2000}).

\bibitem[{\citenamefont{Morigi and Walter}(2001)}]{cooling3}
\bibinfo{author}{\bibfnamefont{G.}~\bibnamefont{Morigi}} \bibnamefont{and}
  \bibinfo{author}{\bibfnamefont{H.}~\bibnamefont{Walter}},
  \bibinfo{journal}{Eur. Phys. J. D} \textbf{\bibinfo{volume}{13}},
  \bibinfo{pages}{261} (\bibinfo{year}{2001}).

\bibitem[{\citenamefont{Wineland and Itano}(1979)}]{doppler}
\bibinfo{author}{\bibfnamefont{D.~J.} \bibnamefont{Wineland}} \bibnamefont{and}
  \bibinfo{author}{\bibfnamefont{W.~M.} \bibnamefont{Itano}},
  \bibinfo{journal}{Phys.\ Rev.\ A} \textbf{\bibinfo{volume}{20}},
  \bibinfo{pages}{1521} (\bibinfo{year}{1979}).

\bibitem[{\citenamefont{Metcalf and van~der Straten}(1999)}]{book2}
\bibinfo{author}{\bibfnamefont{H.~J.} \bibnamefont{Metcalf}} \bibnamefont{and}
  \bibinfo{author}{\bibfnamefont{P.}~\bibnamefont{van~der Straten}},
  \emph{\bibinfo{title}{Laser Cooling and Trapping}}
  (\bibinfo{publisher}{Springer}, \bibinfo{year}{1999}).

\bibitem[{\citenamefont{Monroe et~al.}(1995)}]{cooling5}
\bibinfo{author}{\bibfnamefont{C.}~\bibnamefont{Monroe}} \bibnamefont{et~al.},
  \bibinfo{journal}{Phys.\ Rev.\ Lett.} \textbf{\bibinfo{volume}{75}},
  \bibinfo{pages}{4011} (\bibinfo{year}{1995}).

\bibitem[{\citenamefont{Turchette et~al.}(2000)}]{Heating}
\bibinfo{author}{\bibfnamefont{Q.~A.} \bibnamefont{Turchette}}
  \bibnamefont{et~al.}, \bibinfo{journal}{Phys. Rev. A}
  \textbf{\bibinfo{volume}{61}}, \bibinfo{pages}{063418}
  (\bibinfo{year}{2000}).

\bibitem[{\citenamefont{Raizen et~al.}(1992)\citenamefont{Raizen, Gilligan,
  Berquist, Itano, and Wineland}}]{trap}
\bibinfo{author}{\bibfnamefont{M.~G.} \bibnamefont{Raizen}},
  \bibinfo{author}{\bibfnamefont{J.~M.} \bibnamefont{Gilligan}},
  \bibinfo{author}{\bibfnamefont{J.~C.} \bibnamefont{Berquist}},
  \bibinfo{author}{\bibfnamefont{W.~M.} \bibnamefont{Itano}}, \bibnamefont{and}
  \bibinfo{author}{\bibfnamefont{D.~J.} \bibnamefont{Wineland}},
  \bibinfo{journal}{Phys. Rev. A} \textbf{\bibinfo{volume}{45}},
  \bibinfo{pages}{6493} (\bibinfo{year}{1992}), \bibinfo{note}{and references
  therein.}

\bibitem[{\citenamefont{Berkeland et~al.}(1998)\citenamefont{Berkeland, Miller,
  Bergquist, Itano, and Wineland}}]{Micromotion}
\bibinfo{author}{\bibfnamefont{D.~J.} \bibnamefont{Berkeland}},
  \bibinfo{author}{\bibfnamefont{D.~J.} \bibnamefont{Miller}},
  \bibinfo{author}{\bibfnamefont{J.~C.} \bibnamefont{Bergquist}},
  \bibinfo{author}{\bibfnamefont{W.~M.} \bibnamefont{Itano}}, \bibnamefont{and}
  \bibinfo{author}{\bibfnamefont{D.~J.} \bibnamefont{Wineland}},
  \bibinfo{journal}{J. Appl. Phys.} \textbf{\bibinfo{volume}{83}},
  \bibinfo{pages}{5025} (\bibinfo{year}{1998}).

\bibitem[{\citenamefont{Goldstein}(1981)}]{Goldstein}
\bibinfo{author}{\bibfnamefont{H.}~\bibnamefont{Goldstein}},
  \emph{\bibinfo{title}{Classical Mechanics}}
  (\bibinfo{publisher}{Addison-Wesley}, \bibinfo{year}{1981}),
  \bibinfo{edition}{2nd} ed.

\bibitem[{\citenamefont{Wineland et~al.}(2002{\natexlab{a}})}]{Wineland}
\bibinfo{author}{\bibfnamefont{D.~J.} \bibnamefont{Wineland}}
  \bibnamefont{et~al.} (\bibinfo{year}{2002}{\natexlab{a}}),
  \eprint{quant-ph/0212079}.

\bibitem[{\citenamefont{Wineland
  et~al.}(2002{\natexlab{b}})\citenamefont{Wineland, Bergquist, Bollinger,
  Drullinger, and Itano}}]{added3}
\bibinfo{author}{\bibfnamefont{D.~J.} \bibnamefont{Wineland}},
  \bibinfo{author}{\bibfnamefont{J.~C.} \bibnamefont{Bergquist}},
  \bibinfo{author}{\bibfnamefont{J.~J.} \bibnamefont{Bollinger}},
  \bibinfo{author}{\bibfnamefont{R.~E.} \bibnamefont{Drullinger}},
  \bibnamefont{and} \bibinfo{author}{\bibfnamefont{W.~M.} \bibnamefont{Itano}},
  in \emph{\bibinfo{booktitle}{Proc. 6th Symposium Frequency Standards and
  Metrology}}, edited by \bibinfo{editor}{\bibfnamefont{P.}~\bibnamefont{Gill}}
  (\bibinfo{publisher}{World Scientific}, \bibinfo{address}{Singapore},
  \bibinfo{year}{2002}{\natexlab{b}}), pp. \bibinfo{pages}{361--368}.

\bibitem[{\citenamefont{Cirac et~al.}(1994)\citenamefont{Cirac, Garay, Blatt,
  Parkins, and Zoller}}]{recoil1}
\bibinfo{author}{\bibfnamefont{J.~I.} \bibnamefont{Cirac}},
  \bibinfo{author}{\bibfnamefont{L.~J.} \bibnamefont{Garay}},
  \bibinfo{author}{\bibfnamefont{R.}~\bibnamefont{Blatt}},
  \bibinfo{author}{\bibfnamefont{A.~S.} \bibnamefont{Parkins}},
  \bibnamefont{and} \bibinfo{author}{\bibfnamefont{P.}~\bibnamefont{Zoller}},
  \bibinfo{journal}{Phys.\ Rev. A} \textbf{\bibinfo{volume}{49}},
  \bibinfo{pages}{421} (\bibinfo{year}{1994}).

\bibitem[{\citenamefont{Javanainen and Stenholm}(1980)}]{recoil2}
\bibinfo{author}{\bibfnamefont{J.}~\bibnamefont{Javanainen}} \bibnamefont{and}
  \bibinfo{author}{\bibfnamefont{S.}~\bibnamefont{Stenholm}},
  \bibinfo{journal}{Appl. Phys.} \textbf{\bibinfo{volume}{21}},
  \bibinfo{pages}{35} (\bibinfo{year}{1980}).

\bibitem[{\citenamefont{Dalibard et~al.}(1992)\citenamefont{Dalibard, Raimond,
  and Zinn-Justin}}]{book1}
\bibinfo{editor}{\bibfnamefont{J.}~\bibnamefont{Dalibard}},
  \bibinfo{editor}{\bibfnamefont{J.-M.} \bibnamefont{Raimond}},
  \bibnamefont{and}
  \bibinfo{editor}{\bibfnamefont{J.}~\bibnamefont{Zinn-Justin}}, eds.,
  \emph{\bibinfo{title}{Fundamental systems in quantum optics}}
  (\bibinfo{publisher}{North-Holland}, \bibinfo{year}{1992}),
  \bibinfo{note}{course I}.

\end{thebibliography}

\end{document}